\documentclass[useAMS,usenatbib]{mnras}
\usepackage{graphicx,natbib,amssymb,bm}

\title[Power spectrum of 2MRS]{Angular power spectrum of
galaxies in the 2MASS Redshift Survey}
\author[S. Ando et al.]{Shin'ichiro Ando,$^{1,2}$ Aur{\'e}lien
Benoit-L{\'e}vy,$^{3}$ and Eiichiro Komatsu$^{4,2}$
\\
$^{1}$GRAPPA Institute, University of Amsterdam, 1098XH Amsterdam,
Netherlands\\
$^{2}$Kavli Institute for the Physics and Mathematics of the Universe
(Kavli IPMU, WPI), Todai Institutes for Advanced Study, \\ University
of Tokyo, Kashiwa 277-8583, Japan\\
$^{3}$Sorbonne Universit{\'e}s, UPMC Univ Paris 6 et CNRS, UMR 7095,
Institut d'Astrophysique de Paris, 
75014 Paris, France\\
$^{4}$Max-Planck-Institut f{\"u}r Astrophysik, Karl-Schwarzschild Strasse
1, 85748 Garching, Germany}
\begin{document}

\date{16 June 2017; accepted 6 October 2017}

\maketitle		

\label{firstpage}

\begin{abstract}
We present the measurement and interpretation of the angular
 power spectrum of nearby galaxies in the 2MASS Redshift Survey catalog with
 spectroscopic redshifts up to $z\approx 0.1$. We detect the angular
 power spectrum up to a multipole of $\ell\approx 1000$. We find that
 the measured power spectrum is dominated by galaxies living inside
 nearby galaxy clusters and groups. We use the halo occupation
 distribution (HOD) formalism to model the power spectrum, obtaining a
 fit with reasonable parameters. These HOD parameters are in
 agreement with the 2MASS galaxy distribution we measure toward the
 known nearby galaxy clusters, confirming validity of our analysis.
\end{abstract}


\section{Introduction}
The 2MASS Redshift Survey \citep[2MRS;][]{2MRS} is a spectroscopic
follow-up of galaxies detected in the photometric Two Micron All Sky
Survey \citep[2MASS;][]{2MASS}. The 2MRS provides an excellent probe of
the {\it nearby} distribution of galaxies up to $z\approx 0.1$. Among
many applications of this full-sky map of galaxies, one powerful
application is the cross-correlation with a map of different tracers
with unknown redshifts.

For example, the cross correlation of the photometric 2MASS galaxies
with the cosmic microwave background (CMB) temperature map was used to measure a
dynamical signature of dark energy in the CMB via the integrated
Sachs-Wolfe effect \citep{afshordi}. The cross-correlation with a
gamma-ray map measured by Large Area Telescope (LAT) of the Fermi
satellite has an excellent sensitivity to gamma-rays from annihilation
of dark matter particles \citep{xia,Ando2014,cuoco}. The
cross correlation of the 2MRS catalog with a map of the thermal
Sunyaev-Zeldovich effect derived from the Planck data \citep{plancksz}
has been measured recently \citep{makiya}, and it tells us how hot gas
traces galaxies in the nearby Universe.

To understand properly various such measurements of cross correlations,
we must understand clustering of galaxies in the 2MRS catalog, i.e., 
the auto power spectrum of 2MRS. In this paper, we present the
measurement and interpretation of the 2MRS angular power spectrum.
\citet{Frith2005} presented the angular power spectrum of the 2MASS full
release extended source catalog.

Throughout the paper, we adopt cosmological parameters of
\citet[Table~4, TT+lowP+lensing]{Planck2015}. We define a halo mass ($M
\equiv M_{\rm vir}$) as a mass enclosed within a virial radius $r_{\rm
vir}$, within which the average matter density is $\Delta_{\rm vir}(z)$
times the critical density $\rho_{\rm c}(z)$, where $\Delta_{\rm vir}(z)
\equiv 18\pi^2+82d-39d^2$ with $d \equiv \Omega_{\rm m}(1+z)^3/[\Omega_{\rm
m}(1+z)^3+\Omega_\Lambda]-1$ \citep{Bryan1998}. Conversion from other
definitions of the mass (such as $M_{200}$ defined with radius $r_{200}$
within which the average density is 200 times the critical density) is
done by assuming a Navarro-Frenk-White density profile \citep{NFW} and
using fitting formulae given in \citet{Hu2003}.

\section{Angular power spectrum}
\subsection{Construction of the galaxy density map}
We construct a pixelized count map of the 2MRS galaxies using the
HEALPix\footnote{http://healpix.sourceforge.net} scheme on the sphere,
with the resolution parameter of $N_{\rm{side}}=512$. We then construct
a map of the galaxy density contrast as
\begin{equation}
\delta_{\rm g}= \frac{n_{\rm g}-\bar n_{\rm g}}{\bar n_{\rm g} }\,,
\end{equation}
where $\bar n_{\rm g}$ is the mean number of galaxies per pixel, and $n_{\rm g}$
is the number of galaxies in a given pixel. 

We avoid regions close to the Galactic plane by masking pixels in $|b|
<5^\circ$ for $30^\circ < l < 330^\circ$ and $|b|<10^\circ$ otherwise.
In addition we mask small regions that have very low redshift completeness
as computed by \citet{Lavaux2011} (the white regions in the left panel
of their figure 4). The fraction of sky available for the analysis is
$f_{\rm sky}=0.877$.

\begin{figure}
 \begin{center}
  \includegraphics[width=8cm]{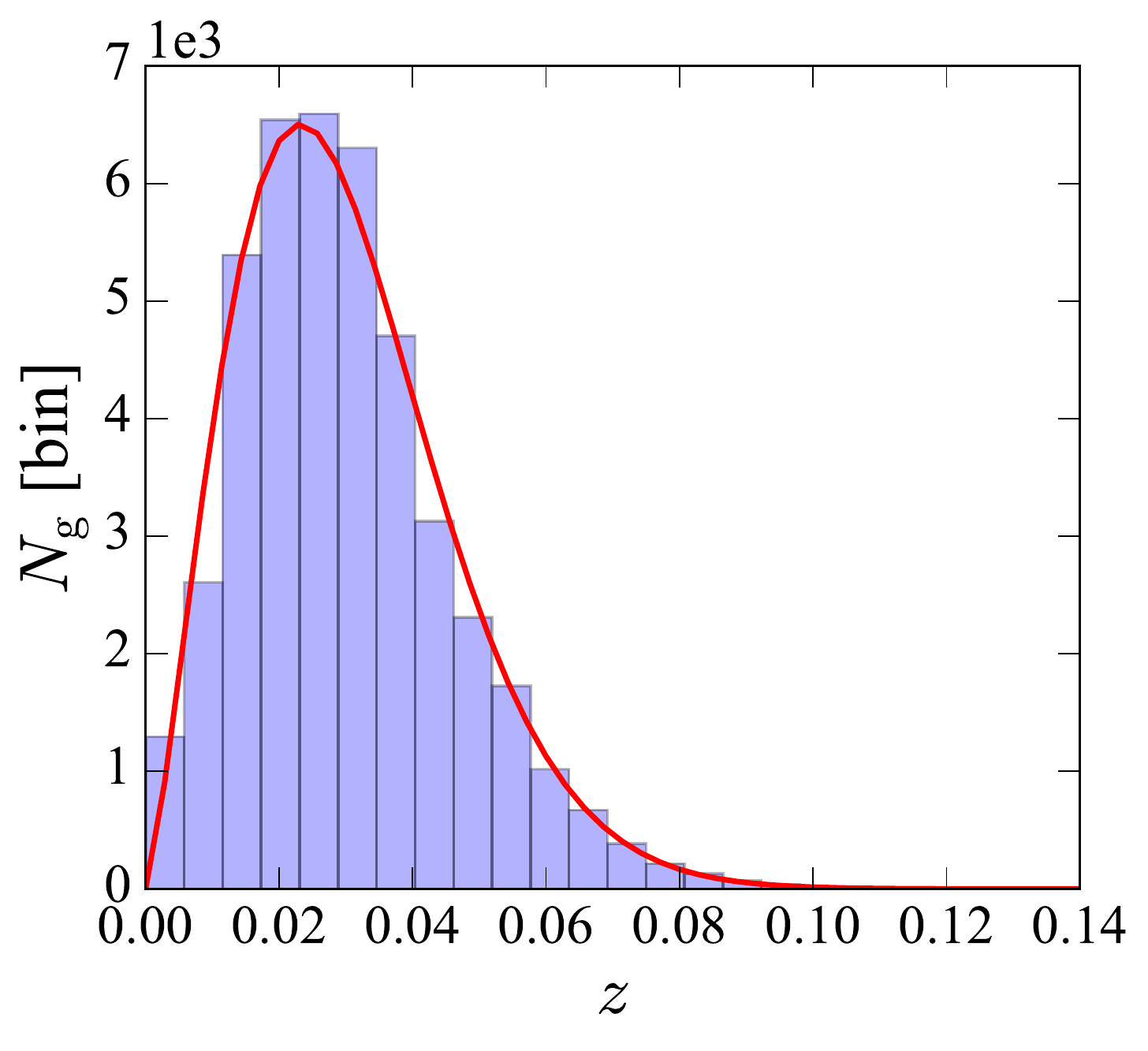}
  \caption{The number of 2MRS galaxies in each redshift bin, compared
  with the fitting functon, equation.~(\ref{eq:dNdz}).}
  \label{fig:Ng_2MRS}
 \end{center}
\end{figure}

The redshift distribution of the 2MRS galaxies is well approximated by
\begin{equation}
 \frac{dN_{\rm g}}{dz} = \frac{N_{\rm g}\beta}{z_0\Gamma[(m+1)/\beta]}
  \left(\frac{z}{z_0}\right)^m
  \exp\left[-\left(\frac{z}{z_0}\right)^{\beta}\right]\,,
  \label{eq:dNdz}
\end{equation}
where $N_{\rm g} = 43182$ is the total number of the galaxies in the 2MRS
catalog in the unmasked pixels. We performed an unbinned likelihood
analysis and found $m = 1.31$, $\beta = 1.64$, and $z_0 = 0.0266$.
Figure~\ref{fig:Ng_2MRS} shows $dN_{\rm g}/dz$ of the 2MRS galaxies.

Redshifts of the 2MRS are claimed to be complete for galaxies up to a
magnitude of $K_s=11.75$, and are nearly volume-limited up to $z\approx
0.02$. Completeness of redshifts as computed by \citet{Lavaux2011} is
quite homogeneous except small regions that have very low
completeness. We mask those regions as described above, and apply no completeness
correction for the derived $\delta_{\rm g}$.

\subsection{Estimation of the angular power spectrum}
\label{sec:cl}
We used the publicly available software
PolSpice\footnote{http://www2.iap.fr/users/hivon/software/PolSpice/} to
estimate the angular power spectrum of $\delta_{\rm g}$. PolSpice allows
for an efficient de-convolution of the harmonic couplings induced by the
mask.

The estimated angular power spectrum, $C_\ell$, is the sum of the signal
and the shot noise, and the latter needs to be subtracted. The shot
noise is usually assumed to be equal to $1/\bar n_{\rm g}$, but this
simple relation may not always hold due to halo exclusion and non-linear
effects \citep{baldauf}. We thus employ here another route to
disentangle signal and noise. We consider a half-sum (HS)
and a half-difference (HD) of the data to obtain directly a noise
estimate. Instead of generating a single density map from a given
catalog, we first divide the catalog into two subsets by randomly
selecting half the galaxies, and construct two galaxy density maps
$\delta_{{\rm g},1}$ and $\delta_{{\rm g},2}$. We then form the
following maps:
\begin{equation}
HS=\frac{\delta_{{\rm g},1}+\delta_{{\rm g,}2}}{2}\,, \quad HD=\frac{\delta_{{\rm g},1}-\delta_{{\rm g},2}}{2}\,,
\end{equation}
As the division of the catalog into subsets is a random process, the
estimated $HD$ changes slightly depending on realizations. These
changes do not affect our conclusion, but we shall explore the impacts
of randomness on the best-fitting model parameters in
Sec.~\ref{sec:trispectrum}.

The power spectrum extracted from the $HS$ map contains both the signal
and noise, whereas that from the $HD$ map contains only the noise. Hence
our estimator for the angular power spectrum is given by
\begin{equation}
\hat C_\ell =   C_\ell^{HS} - C_\ell^{HD}\,.
\end{equation}

\begin{figure}
 \begin{center}
  \includegraphics[width=8cm]{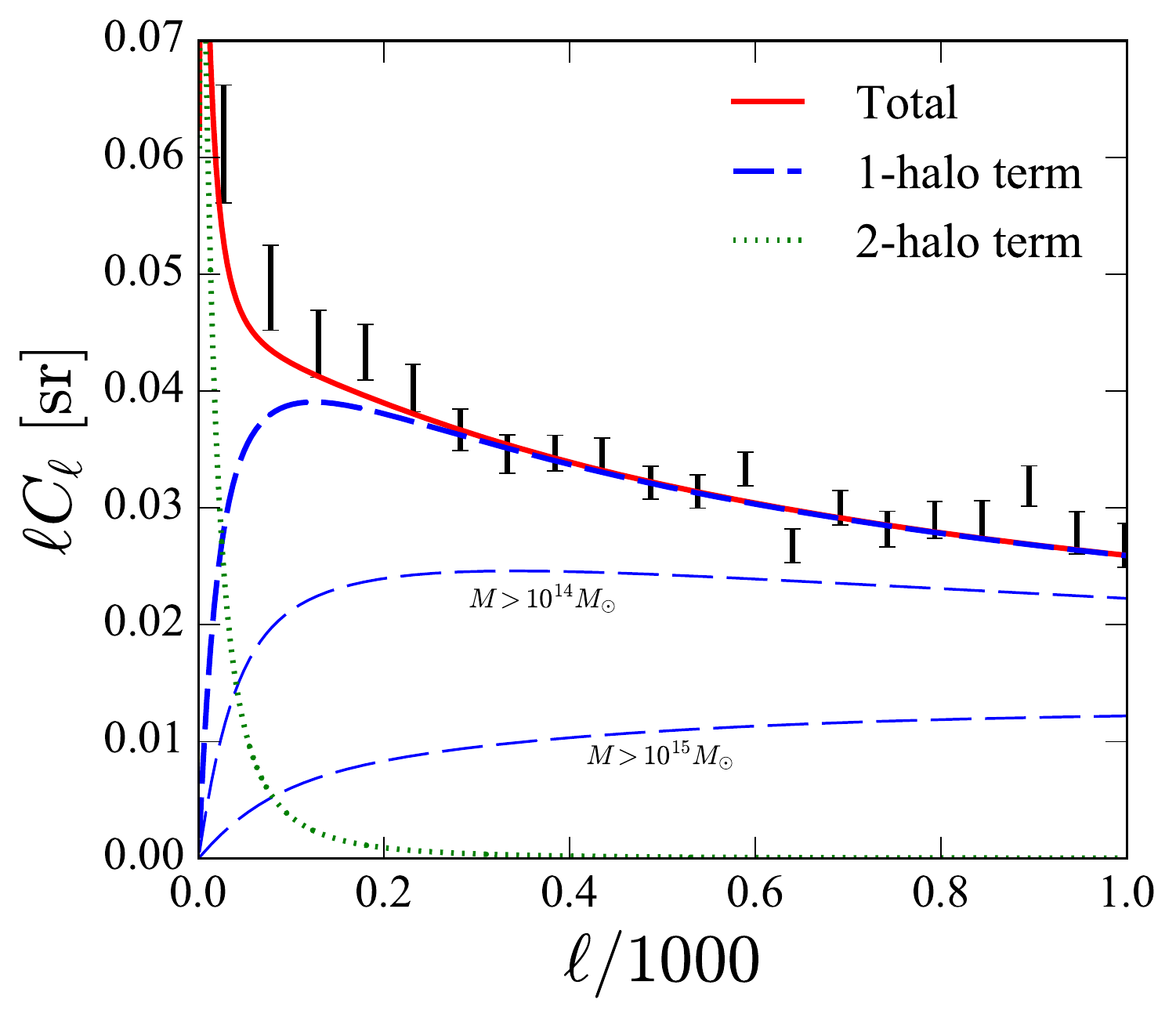}
  \includegraphics[width=8cm]{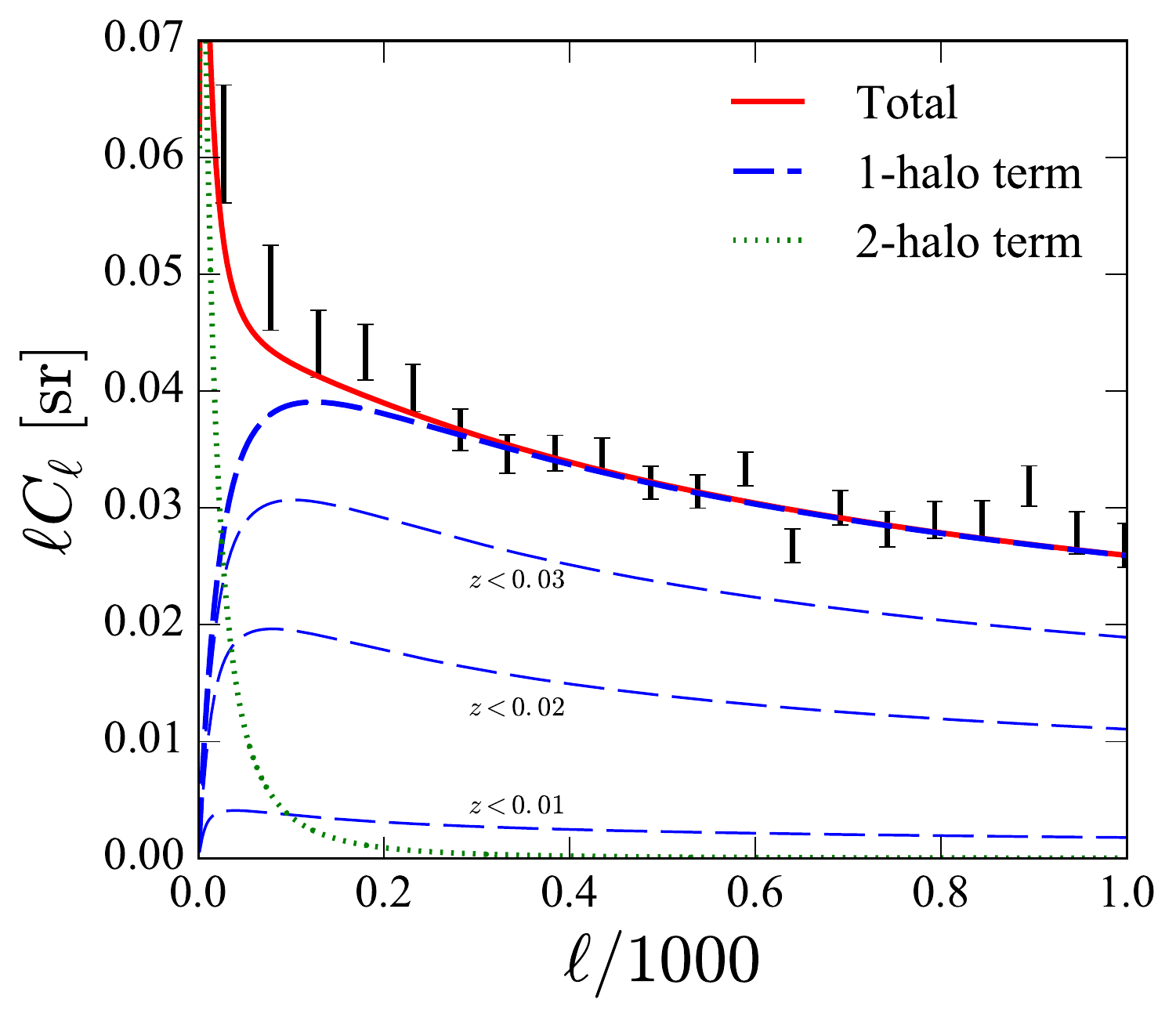}
  \caption{Angular power spectrum of the 2MRS galaxies, compared
  with the best-fitting model (solid; see Sec.~\ref{sec:model}). The
  one- and two-halo terms are shown as the thick dashed and the dotted curves,
  respectively. The thin dashed curves show contributions from mass
  ranges relevant for clusters of galaxies as indicated (top), and
  those from a few different redshift ranges (bottom). The error bars
  show the diagonal elements of the covariance matrix (see
  Sec.~\ref{sec:errorbar}).}
  \label{fig:Cl}
 \end{center}
\end{figure}

We present the measured power spectrum minus the noise power in
figure~\ref{fig:Cl}. We detect the power spectrum clearly all the way to
the maximum multipole reliably resolved by the HEALPix map with
$N_{\rm{side}}=512$, i.e., $\ell_{\rm max}=2N_{\rm{side}}=1024$. This
indicates that the 2MRS galaxies are strongly clustered: we have
43182 galaxies over 87.7\% of the sky, and thus the mean galaxy number
density is only 1.2~deg$^{-2}$. As we show in this paper, many of the
2MRS galaxies reside in galaxy clusters and groups, producing power at
multipoles much higher than that of the mean separation of galaxies.

We tabulate the measured power spectrum data with diagonal elements of
the covariance matrix in Table~\ref{tab:cl}.

\begin{table*}
  \begin{center}
   \caption{Power spectrum of the 2MRS, the Gaussian variance, and the
   total variance.}
   \label{tab:cl}
 \begin{tabular}{ccrrr} \hline
  Mean multipole & Multipole range & $\ell C_\ell$ [10$^{-2}$~sr] &
	      $\sqrt{\mbox{Cov}(\ell C_\ell,\ell C_\ell)_{\rm Gauss}}$
  [10$^{-2}$~sr] &
  $\sqrt{\mbox{Cov}(\ell C_\ell,\ell C_\ell)}$  [10$^{-2}$~sr] \\ \hline
$27$ & $2 \le \ell <53$ & $6.11$ & $0.32$ & $0.50$ \\
$78$ & $53 \le \ell <104$ & $4.88$ & $0.12$ & $0.36$ \\
$129$ & $104 \le \ell <155$ & $4.40$ & $0.10$ & $0.29$ \\
$180$ & $155 \le \ell <206$ & $4.33$ & $0.10$ & $0.24$ \\
$231$ & $206 \le \ell <257$ & $4.03$ & $0.10$ & $0.20$ \\
$282$ & $257 \le \ell <308$ & $3.67$ & $0.10$ & $0.18$ \\
$333$ & $308 \le \ell <359$ & $3.46$ & $0.10$ & $0.16$ \\
$384$ & $359 \le \ell <410$ & $3.47$ & $0.11$ & $0.15$ \\
$435$ & $410 \le \ell <461$ & $3.45$ & $0.11$ & $0.15$ \\
$487$ & $461 \le \ell <513$ & $3.22$ & $0.11$ & $0.14$ \\
$538$ & $513 \le \ell <564$ & $3.14$ & $0.12$ & $0.14$ \\
$589$ & $564 \le \ell <615$ & $3.33$ & $0.12$ & $0.14$ \\
$640$ & $615 \le \ell <666$ & $2.68$ & $0.13$ & $0.14$ \\
$691$ & $666 \le \ell <717$ & $3.00$ & $0.14$ & $0.15$ \\
$742$ & $717 \le \ell <768$ & $2.82$ & $0.14$ & $0.15$ \\
$793$ & $768 \le \ell <819$ & $2.90$ & $0.15$ & $0.16$ \\
$844$ & $819 \le \ell <870$ & $2.90$ & $0.16$ & $0.16$ \\
$895$ & $870 \le \ell <921$ & $3.19$ & $0.17$ & $0.17$ \\
$946$ & $921 \le \ell <972$ & $2.79$ & $0.18$ & $0.18$ \\
$998$ & $972 \le \ell <1024$ & $2.68$ & $0.18$ & $0.19$ \\
\hline
 \end{tabular}
  \end{center}
\end{table*}

\subsection{Covariance matrix}
\label{sec:errorbar}
To estimate the covariance matrix of the power spectrum,
$\mbox{Cov}(C_\ell,C_{\ell'})\equiv
\langle(C_\ell-\langle C_\ell\rangle)(C_{\ell'}-\langle
C_{\ell'}\rangle)\rangle$
where $\langle\dots\rangle$ denotes the ensemble
average, we take a two-step approach. We first start with a
diagonal Gaussian covariance given by
\begin{equation}
\mbox{Cov}(C_\ell, C_{\ell^\prime})_{\rm Gauss} =
 \frac{2{(C_\ell^{tot})}^2 \delta_{\ell \ell^\prime}}{f_{\rm sky} (2\ell+1)\Delta\ell}\,,
\end{equation}
where $\delta_{\ell \ell^\prime}$ is the Kronecker delta, $\Delta\ell$
is the size of multipole bins,
$f_{\rm sky}$ is the sky fraction used in the analysis, and
$C_\ell^{tot}$ is the sum of the signal and the noise power, i.e.,
$C_\ell^{tot}= C_\ell^{HS}$. To reduce an artificial scatter in
the covariance matrix, we smooth the measured total power spectrum
before computing the covariance matrix.

To this Gaussian covariance, we add a non-Gaussian contribution
from the connected part of the trispectrum, $T_{\ell\ell'}$, as \citep[e.g.,][]{KomatsuSeljak}
\begin{equation}
\mbox{Cov}(C_\ell,C_{\ell'}) = \frac{1}{f_{\rm
 sky}}
 \left[\frac{2(C_\ell^{tot})^2\delta_{\ell\ell'}}{(2\ell+1)\Delta\ell}+\frac{T_{\ell\ell'}}{4\pi}\right]\,.
\end{equation}
While the trispectrum term is usually ignored in the galaxy power
spectrum analysis, it is important for the 2MRS angular power spectrum
because the power spectrum is dominated by galaxies living inside nearby
clusters and groups. One may understand this intuitively by noting that
the presence of a few nearby massive clusters adds power at all
multipoles (see Sec.~\ref{sec:individual}) and thus different multipole
bins become strongly correlated. This is the sign that the trispectrum term is
important.

We shall describe details of the computation of $T_{\ell\ell'}$ in
Sec.~\ref{sec:trispectrum}, and only outline our procedure here. As the
trispectrum term depends on the model of the power spectrum itself, we cannot
calculate $T_{\ell\ell'}$ {\it a priori}. Thus, we first determine the
model parameters using the Gaussian covariance in the likelihood analysis. We
then calculate $T_{\ell\ell'}$ for the best-fitting model parameters,
and re-determine the model parameters using the sum of the Gaussian term
and $T_{\ell\ell'}$ in the likelihood. The error bars shown in
figure~\ref{fig:Cl} are the diagonal elements of the total covariance
matrix including the trispectrum term.

The square-root of the diagonal
elements of the Gaussian and total covariance matrices are reported in
Table~\ref{tab:cl}.

\section{Model}
\label{sec:model}
\subsection{Halo model}
We use a halo model \citep{Seljak2000} to compute the model galaxy power
spectrum. We assume that galaxies reside in virialized dark matter halos.
In this framework, the galaxy power spectrum is divided into
one-halo (1h) and two-halo (2h) terms as
$P_{\rm g}(k,z) = P_{\rm g}^{\rm 1h}(k,z) + P_{\rm g}^{\rm 2h}(k,z)$, where
\begin{eqnarray}
 P_{\rm g}^{\rm 1h}(k,z) &=& \frac{1}{\langle n_{\rm
  g}(z)\rangle^2} \int dM \frac{dn}{dM} \left[2\langle N_{\rm
					 sat}|M\rangle \tilde u_{\rm
					 sat}(k,M)
  \right.\nonumber\\&&{}\left. + \langle N_{\rm sat}|M\rangle^2 |\tilde
  u_{\rm sat}(k,M)|^2\right]\,, \\
 P_{\rm g}^{\rm 2h}(k,z) &=& b_{\rm g}^2(k,z) P_{\rm lin}(k,z)\,.
\end{eqnarray}
Here $dn/dM$ is the halo mass function for which we adopt
a model of \citet{Tinker2008, Tinker2010}, $\tilde u_{\rm sat}(k,M)$ is
the Fourier transform of the radial distribution of satellite galaxies
normalised such that $\tilde u_{\rm sat}(k,M)\to 1$ as $k\to 0$, 
$P_{\rm lin}(k,z)$ is the linear matter power spectrum computed with
{\tt CLASS} \citep{CLASS}, $\langle n_{\rm g}(z) \rangle$ is the mean number
density of galaxies given by
\begin{equation}
  \langle n_{\rm g}(z)\rangle =
  \int dM \frac{dn}{dM} \left(\langle N_{\rm cen}|M\rangle
     + \langle N_{\rm sat}|M\rangle\right),
  \label{eq:nHOD}
\end{equation}
and $b_{\rm g}(k,z)$ is the scale-dependent galaxy bias given by
\begin{eqnarray}
  b_{\rm g}(k,z) &=& \frac{1}{\langle n_{\rm g}(z)\rangle}
  \int dM \frac{dn}{dM} \left[\langle N_{\rm cen}|M\rangle
    \right.\nonumber\\&&{}\left.
     + \langle N_{\rm sat}|M\rangle \tilde u_{\rm sat}(k|M)\right]
    b_1(M,z),
\end{eqnarray}
 with a linear halo bias $b_1(M,z)$ of \citet{Tinker2010}.
The other functions, $\langle N_{\rm cen}|M\rangle$ and $\langle N_{\rm
sat}|M\rangle$, are called the halo occupation distribution (HOD), and
we shall describe them in Sec.~\ref{sec:hod}.

\begin{figure*}
 \begin{center}
  \includegraphics[width=15cm]{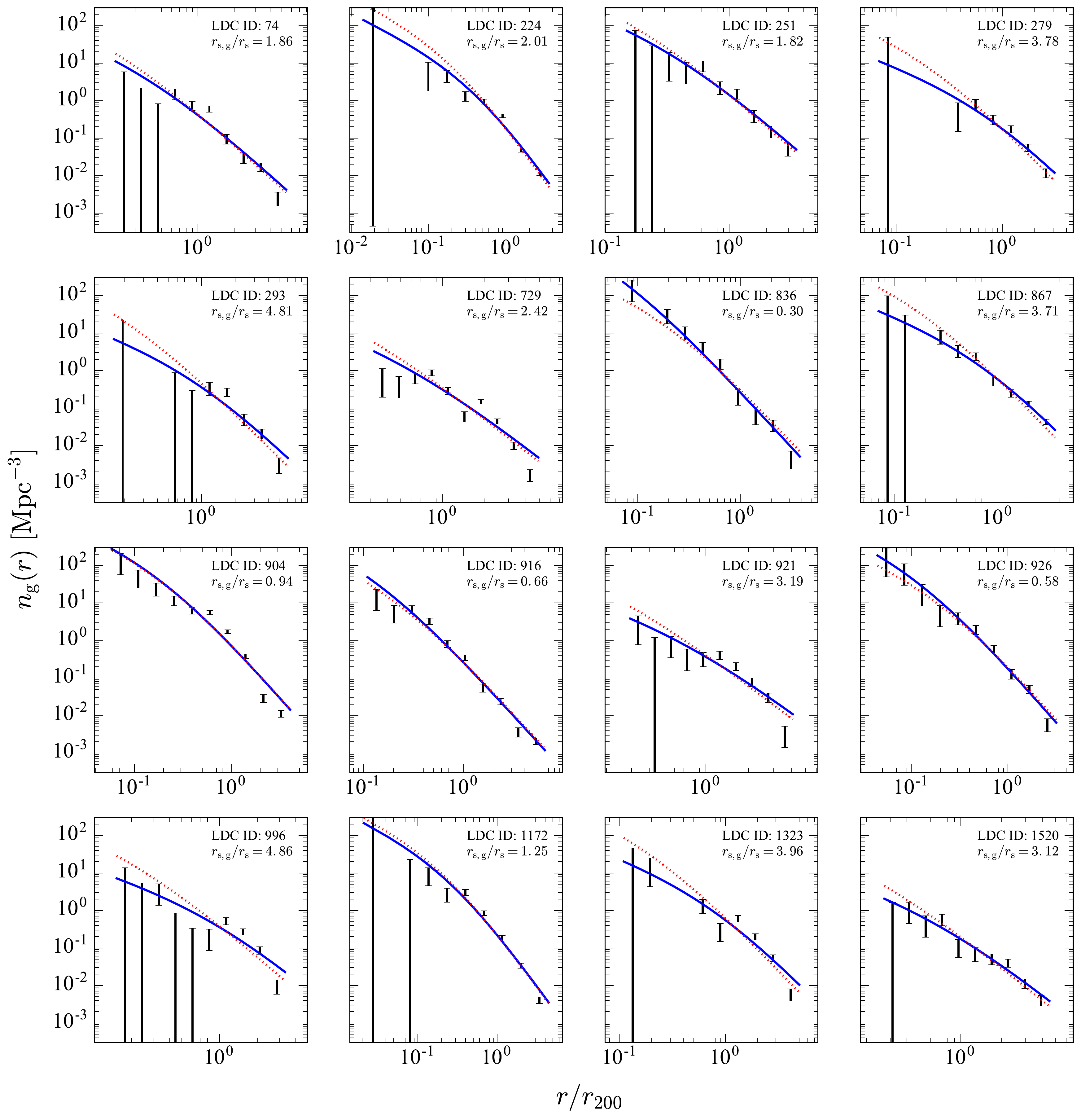}
  \caption{Number density profiles of member galaxies in the 16 largest galaxy
  clusters (with ID numbers) found in the LDC catalog of 2MRS galaxies
  \citep{Crook2007, Crook2008}. The solid and dotted curves show NFW
  profiles with the best-fitting $r_{\rm s,g}/r_{\rm s}$ and with a
  fixed $r_{\rm s,g}/r_{\rm s} = 1$, respectively.}
  \label{fig:gal_prof_clusters}
 \end{center}
\end{figure*}

We assume that the radial distribution of satellite galaxies in a halo
follows an NFW profile \citep{NFW} characterized by a scale
radius for galaxies $r_{\rm s,g}$ up to some maximum radius $r_{\rm max,g}$:
\begin{equation}
 u_{\rm sat}(r) \propto \frac{1}{(r/r_{\rm s,g})(r/r_{\rm s,g}+1)^2}
  \Theta(r_{\rm max,g}-r)\,,
\end{equation}
where $\Theta$ is the Heaviside step function.
In figure~\ref{fig:gal_prof_clusters}, we show radial distribution
profiles of member galaxies in the 16 largest galaxy clusters found in the
low-density-contrast (LDC) catalog of 2MRS \citep{Crook2007, Crook2008}.
We find that satellite galaxies are distributed out to 3--6
times $r_{200}$.\footnote{The maximum radius $r_{\rm max,g}$ depends on how
one defines galaxy clusters.
In a more recent catalog of \citet{Tully2015}, clusters are defined to
be more compact than the LDC definition of \citet{Crook2007,
Crook2008}, yielding smaller values of $r_{\rm max, g} / r_{200}$.
In order to accommodate this ambiguity, we shall treat $r_{\rm
max,g} / r_{200}$ as a free parameter.}

The best-fitting values for the ratio of the scale
radius of galaxies to that of dark matter $r_{\rm g,s} / r_{\rm s}$
vary across clusters. We use the mass-concentration relation of
\citet{SCP2014} to compute $r_{\rm s}$ as a function of halo masses,
which are inferred from dynamical analysis of the member
galaxies in each cluster \citep{Crook2007, Crook2008}.
The solid curves in figure~\ref{fig:gal_prof_clusters} show NFW
profiles with the best-fitting $r_{\rm s,g}/r_{\rm s}$ from unbinned
likelihood analysis. Since the parameter $r_{\rm s,g} / r_{\rm s}$ varies
from cluster to cluster, and the cluster mass estimates as well as
intrinsic scatter in mass-concentration relation are uncertain, we shall
treat it as a free parameter.

The angular power spectrum is obtained by projecting $P_{\rm
g}(k,z)$ onto two-dimensional sky.
Using Limber's approximation, we obtain
\begin{equation}
 C_\ell^{\rm g} = \int \frac{d\chi}{\chi^2} W_{\rm g}^2(z)
  P_{\rm g}\left(\frac{\ell}{\chi},z\right)\,,
\end{equation}
where  $\chi$ is the comoving distance to a galaxy at a redshift $z$,
and the function $W_{\rm g} \equiv (d\ln N_{\rm g} /dz) (d\chi /
dz)^{-1}$ is computed with equation~(\ref{eq:dNdz}).

\subsection{Halo Occupation Distribution}
\label{sec:hod}
Following \citet{Zheng2005}, we parameterize the HOD functions
determining the mean number of central galaxies, $\langle N_{\rm
cen}|M\rangle$, and that of satellite galaxies, $\langle N_{\rm
sat}|M\rangle$, in a dark matter halo of mass $M$ as
\begin{eqnarray}
 \langle N_{\rm cen}|M\rangle &=& \frac{1}{2}
  \left[1+\mbox{erf}\left(\frac{\log M - \log M_{\rm min}}{\sigma_{\log
		     M}}\right)\right]\,,\\
  \langle N_{\rm sat}|M\rangle &=& \left(\frac{M-M_{0}}{M_1}\right)^\alpha
  \Theta(M-M_{0})\,,
\end{eqnarray}
where erf is the error function. We fix $\sigma_{\log M} = 0.15$ and
$M_0 = M_{\rm min}$. The latter assumption ensures that satellite
galaxies form only in halos that host a central galaxy. We therefore
have three HOD parameters $\{M_{\rm min}, M_1, \alpha\}$.

We do not vary the HOD parameters as a function of $z$. In principle
this approximation may break down for a magnitude-limited sample
such as the 2MRS, as the dominant galaxy population may change depending
on $z$. As the 2MRS is nearly volume-limited up to $z\approx 0.02$, it
is safe to use the HOD up to that redshift; however, we need to be
careful when interpreting the HOD parameters derived from the whole
sample. We shall come back to this point later.

\section{Results and interpretation}
\subsection{Posterior distribution of the model parameters}
\label{sec:trispectrum}
We use Bayes' theorem to compute the posterior distribution of the model
parameters $\bm\vartheta$ given the data $\bm d$,
$P(\bm\vartheta|\bm d)$,
with $\bm\vartheta = (r_{\rm max,g} / r_{200}, r_{\rm
s,g}/r_{\rm s}, \log M_{\rm min}, \log M_{1}, \alpha)$ from the
likelihood of the data given a model, $\mathcal L(\bm d|\bm\vartheta)$:
\begin{equation}
 P(\bm\vartheta|\bm d) \propto P(\bm\vartheta) \mathcal L(\bm
  d|\bm\vartheta)\,.
\end{equation}
We adopt flat priors on the model parameters, i.e., $P(\bm\vartheta)={\rm
constant}$ within the parameter ranges given in the second row of
Table~\ref{table:posterior} and $P(\bm\vartheta)=0$ otherwise.

\begin{table*}
  \begin{center}
   \caption{Medians and $1\sigma$ credible intervals of the posterior
   distributions of the model parameters, for three
   independent runs of computing the angular power spectrum. The second
   row shows the prior range, within which each parameter was drawn from
   a flat distribution. The last column shows $\chi_{\rm min}^2$ compared
   with a degree of freedom (dof).}
   \label{table:posterior}
 \begin{tabular}{rccccccc} \hline
  Catalog & $r_{\rm max,g}/r_{200}$ & $r_{\rm s,g}/r_{\rm s}$ & $\log(M_{\rm min}/M_{\odot})$ &
  $\log(M_{1}/M_{\odot})$& $\alpha$ & $\chi_{\rm min}^2 / {\rm dof}$ \\ \hline
  Prior & 0--15 & 0--15 & 9--14 & 9--14 & 0--2 \\ \hline
  Run 1 & $6.9^{+3.0}_{-1.7}$ & $0.62^{+0.13}_{-0.12}$ &
	      $11.84^{+0.17}_{-0.14}$ & $11.98^{+0.18}_{-0.24}$ &
		      $0.849^{+0.073}_{-0.088}$ & $33.3/16$\\
  Run 2 & $6.4^{+3.1}_{-1.6}$ & $0.72^{+0.15}_{-0.14}$ &
	      $11.87^{+0.19}_{-0.17}$ & 
		  $11.97^{+0.21}_{-0.27}$ & $0.855^{+0.087}_{-0.096}$ & $29.7/16$\\
  Run 3 & $6.7^{+2.5}_{-1.5}$ & $0.61^{+0.11}_{-0.10}$ &
	      $11.83^{+0.15}_{-0.13}$ &
		  $11.97^{+0.17}_{-0.21}$ & $0.837^{+0.068}_{-0.074}$ &
			  $47.2/16$ \\
\hline
 \end{tabular}
  \end{center}
\end{table*}

As for the data $\bm d$, we use the angular power spectrum of 2MRS
galaxies $C_\ell^{\rm 2MRS}$, as well as the number of 2MRS galaxies up
to $z = 0.01$, $N_{\rm 2MRS}^{z<0.01} = 3200$, in which the sample is
safely volume-limited. We approximate the likelihood of the data as a Gaussian:
\begin{eqnarray}
 -2\ln\mathcal L(\bm d|\bm\vartheta) &=& \sum_{\ell\ell'}
  \left[C_{\ell}^{\rm th}(\bm\vartheta)-C_{\ell}^{\rm
   2MRS}\right]\mbox{Cov}(C_\ell,C_{\ell^\prime})^{-1}
   \nonumber\\&&\times{}
   \left[C_{\ell^\prime}^{\rm th}(\bm\vartheta)-C_{\ell'}^{\rm
   2MRS}\right]
  \nonumber\\&&{}
  + \frac{\left[N_{\rm th}^{z<0.01}(\bm\vartheta)-N_{\rm
	   2MRS}^{z<0.01}\right]^2}{N_{\rm 2MRS}^{z<0.01}}\,.
\end{eqnarray}
Here the quantities with `th' represent the model predictions given
$\bm\vartheta$, and
$\mbox{Cov}(C_\ell,C_{\ell'})$ is the covariance matrix.

We use the Markov-Chain Monte Carlo (MCMC) to explore the parameter
space. To this end we use the {\tt MultiNest}
\citep{MultiNest1, MultiNest2, MultiNest3} package for the MCMC and
obtain the posterior distributions of the model parameters.

As described in Sec.~\ref{sec:errorbar}, we first run the MCMC with the
Gaussian covariance matrix, and extract the best-fitting model parameters. We
then use those parameters to calculate the trispectrum as
\begin{eqnarray}
T_{\ell\ell'} &=& \int d\chi \frac{W_{\rm g}^4(z)}{\chi^6 \langle n_{\rm g}(z)\rangle^4} \int dM \frac{dn}{dM}
\nonumber\\ && {}\times
\left[2\langle N_{\rm sat}|M\rangle \tilde u_{\rm sat}\left(\frac{\ell}{\chi},M\right) 
\right.\nonumber\\ && {}\left.
+ \langle N_{\rm sat}|M\rangle^2 \left|\tilde u_{\rm sat}\left(\frac{\ell}{\chi},M\right)\right|^2\right]
\nonumber\\ && {}\times
\left[2\langle N_{\rm sat}|M\rangle \tilde u_{\rm sat}\left(\frac{\ell'}{\chi},M\right) 
\right.\nonumber\\ && {}\left.
+ \langle N_{\rm sat}|M\rangle^2 \left|\tilde u_{\rm sat}\left(\frac{\ell'}{\chi},M\right)\right|^2\right]\,.
\end{eqnarray}
We run the MCMC again to obtain the final results. We find good
convergence of the MCMC chains.

\begin{figure*}
 \begin{center}
  \includegraphics[width=16cm]{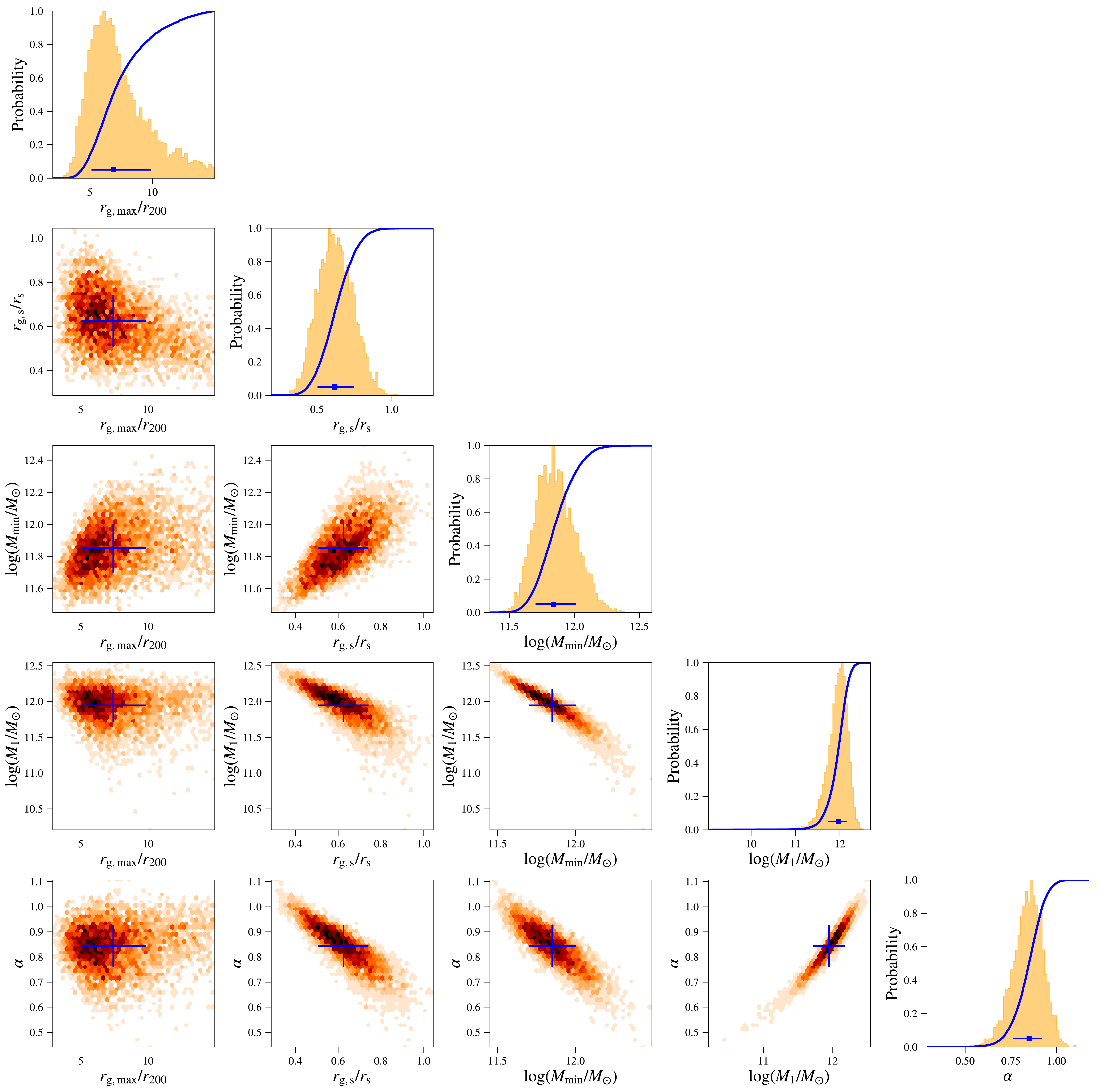}
  \caption{Posterior distribution of the model parameters. 
  The diagonal panels show the one-dimensional distribution
  marginalised over the other parameters as well as the cumulative
  distribution (line). The other panels show the two-dimensional marginalised
  joint distributions. The cross symbols show the mean and the 1$\sigma$
  uncertainty.}
  \label{fig:2mrs_1175_smallmask_run1pwf_iter2_marg}
 \end{center}
\end{figure*}

Since we estimate the noise contribution to the power spectrum by
randomly dividing the 2MRS catalog into two (see Sec.~\ref{sec:cl}), the
galaxy density field used for computing the power spectrum contains some
randomness. To quantify the impacts of this randomness, we calculated
three sets of density fields with different initial random seeds, and
repeated the parameter estimation. The power spectrum shown in
figure~\ref{fig:Cl} is the first set ``run 1'', whose posterior
distribution of the parameters is shown in
figure~\ref{fig:2mrs_1175_smallmask_run1pwf_iter2_marg}. The posterior
distributions for the ``run 2'' and ``run 3'' are similar.
The parameter values are summarized in Table~\ref{table:posterior}. We find
that all runs give similar results.

In the top panel of figure~\ref{fig:Cl}, we show the contributions from
different ranges of halo masses. We find that the angular power spectrum
is dominated by the one-halo term at $\ell \gtrsim 30$. In other words,
the 2MRS angular power spectrum is dominated by galaxies residing in
nearby galaxy clusters and groups.

The derived HOD parameters (Table~\ref{table:posterior}) are reasonable:
the galaxy distribution extends out to several times $r_{200}$; the
galaxy and dark matter scale radii are similar on average, with a slight
preference for $r_{\rm s,g}/r_{\rm s}<1$, though this parameter varies
significantly across individual halos (see
figure~\ref{fig:gal_prof_clusters}); the mass scales enter into the HOD
are on the order of the Milky Way mass; and the number of satellite
galaxies increases with the host halo mass slightly slower than a linear
relation.

The best-fitting models give $\chi_{\rm min}^2 = -2 \ln \mathcal L_{\rm
max}=33.3$, 29.7, and 47.2 for 16 degrees of freedom (i.e., 21 data
points and 5 parameters) for the run 1 through 3. Taken at the face
values, the probabilities to exceed (PTE) these $\chi^2_{\rm min}$ are
$6.7\times 10^{-3}$, $2.0\times 10^{-2}$, and $6.3\times 10^{-5}$. These
PTE are rather low except for the run 2. This may suggest that our HOD
modeling is too simplistic, requiring more sophisticated ones
instead. For example, as we stated in Sec.~\ref{sec:hod}, the use of the
HOD model for a magnitude-limited sample would introduce an error. (The
2MRS is nearly volume-limited up to $z\approx 0.02$.) Such modeling
error would not only affect the model power spectrum, but also the
covariance matrix via the trispectrum term, potentially changing a
$\chi^2$ value. Nonetheless, the model presented here will probably be
good enough for interpreting various cross correlation results.

\subsection{Contribution from the known galaxy groups and clusters}
\label{sec:individual}
Figure~\ref{fig:Cl} shows contributions from the mass ranges relevant
for clusters of galaxies: $M > 10^{14}~M_\odot$ and $>
10^{15}~M_\odot$. We find that most of the power at small angular scales
can be attributed to galaxies residing in galaxy clusters. We can test
this result directly by summing up the contributions from galaxy groups
and clusters we see in the 2MRS. We use recent group catalogs of 2MRS
by \cite{Tully2015} and \citet{Lu2016}.

\begin{figure}
 \begin{center}
  \includegraphics[width=8cm]{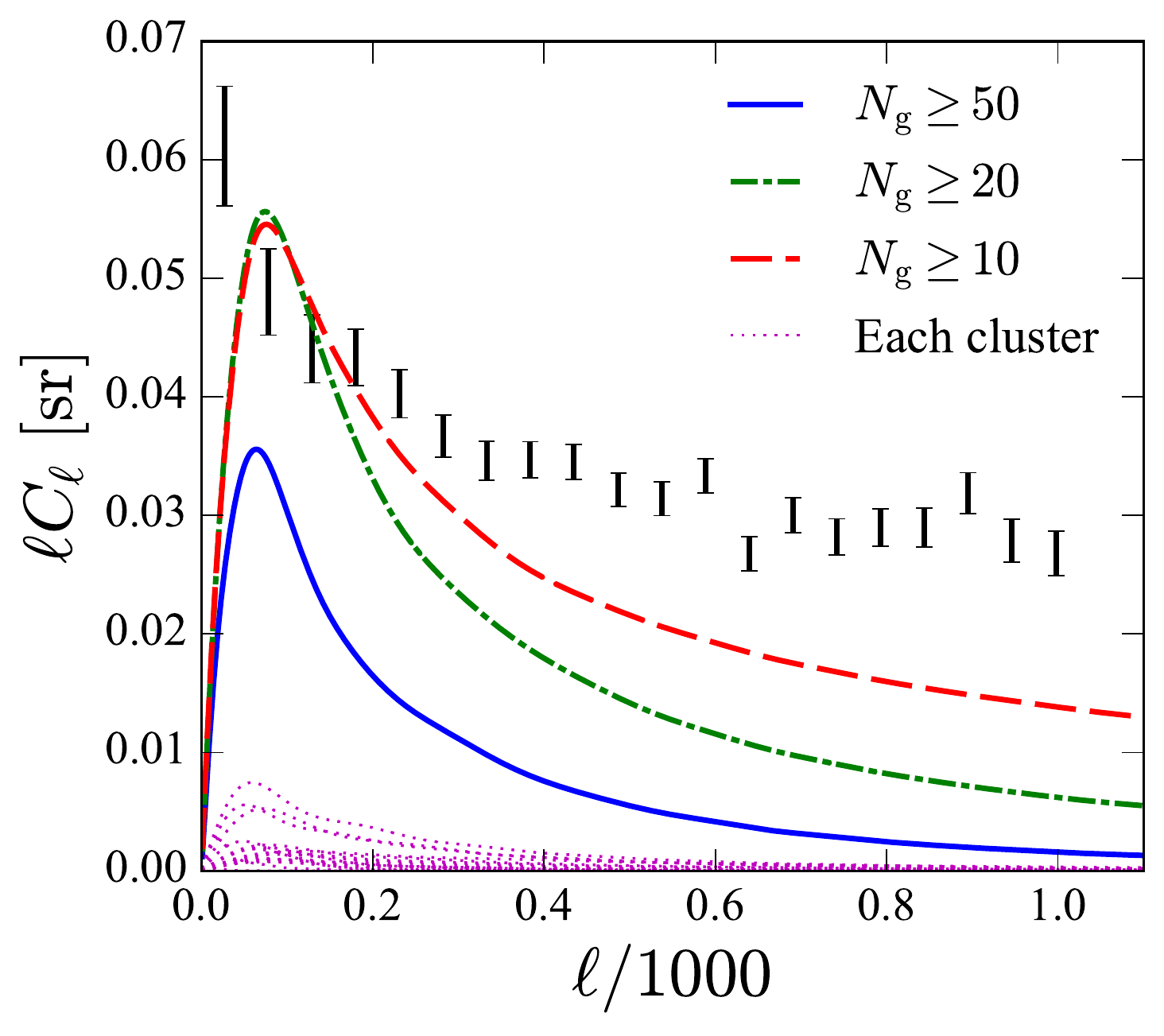}
  \caption{Angular power spectrum of galaxies residing in groups and
  clusters detected 
  in the 2MRS. The cumulative contributions from groups and clusters
  above a certain size are shown as the solid ($N_{\rm g}\ge 50$), dot-dashed
  ($N_{\rm g}\ge 20$), and dashed ($N_{\rm g} \ge 10$) curves. 
  There are 17, 78, and 299 groups and clusters containing 1578, 3376,
  and 6217 galaxies in these samples respectively. The dotted
  curves show $C_\ell$ from each of the 17 largest clusters.}
  \label{fig:Cl_cluster}
 \end{center}
\end{figure}

We compute a contribution to the angular power spectrum from each
of the groups and clusters as
\begin{equation}
 C_\ell^{\rm c} = \frac{4\pi}{N_{\rm 2MRS}^2}
  \left(N_{\rm g}^{\rm c}\right)^2 \left|\tilde u_{\rm g}^{\rm c}
   \left(\frac{\ell}{\chi_{\rm c}},M_{\rm c}\right)\right|^2\,,
  \label{eq:Cl_cluster}
\end{equation}
where ``c'' denotes a group or a cluster, and $\tilde
u_{\rm g}^{\rm c}$ is the Fourier transform of the NFW profile
with the best-fitting $r_{\rm s,g}$ and $r_{\rm max, g}$ for
individual groups and clusters.
Figure~\ref{fig:Cl_cluster} shows contributions to $C_\ell$ from
the catalogs with various numbers of member galaxies, $N_{\rm g}\ge 50$,
$\ge 20,$ and $\ge 10$, in which there are 17, 78, and 299 groups and
clusters, respectively.

We also show individual contributions from the 17 largest clusters.
We find that about ten per cent of the power in sub-degree angular scales
comes from these 17 clusters that host more than 50 member galaxies. The
remaining power comes from hundreds of galaxy groups and clusters.
For smaller groups, the approximation used in
equation~(\ref{eq:Cl_cluster}) becomes less valid, because one can no
longer regard the galaxy distribution as given by the smooth NFW
function. Nonetheless figure~\ref{fig:Cl_cluster} provides a useful
cross-check of our result.

\begin{figure}
 \begin{center}
  \includegraphics[width=8cm]{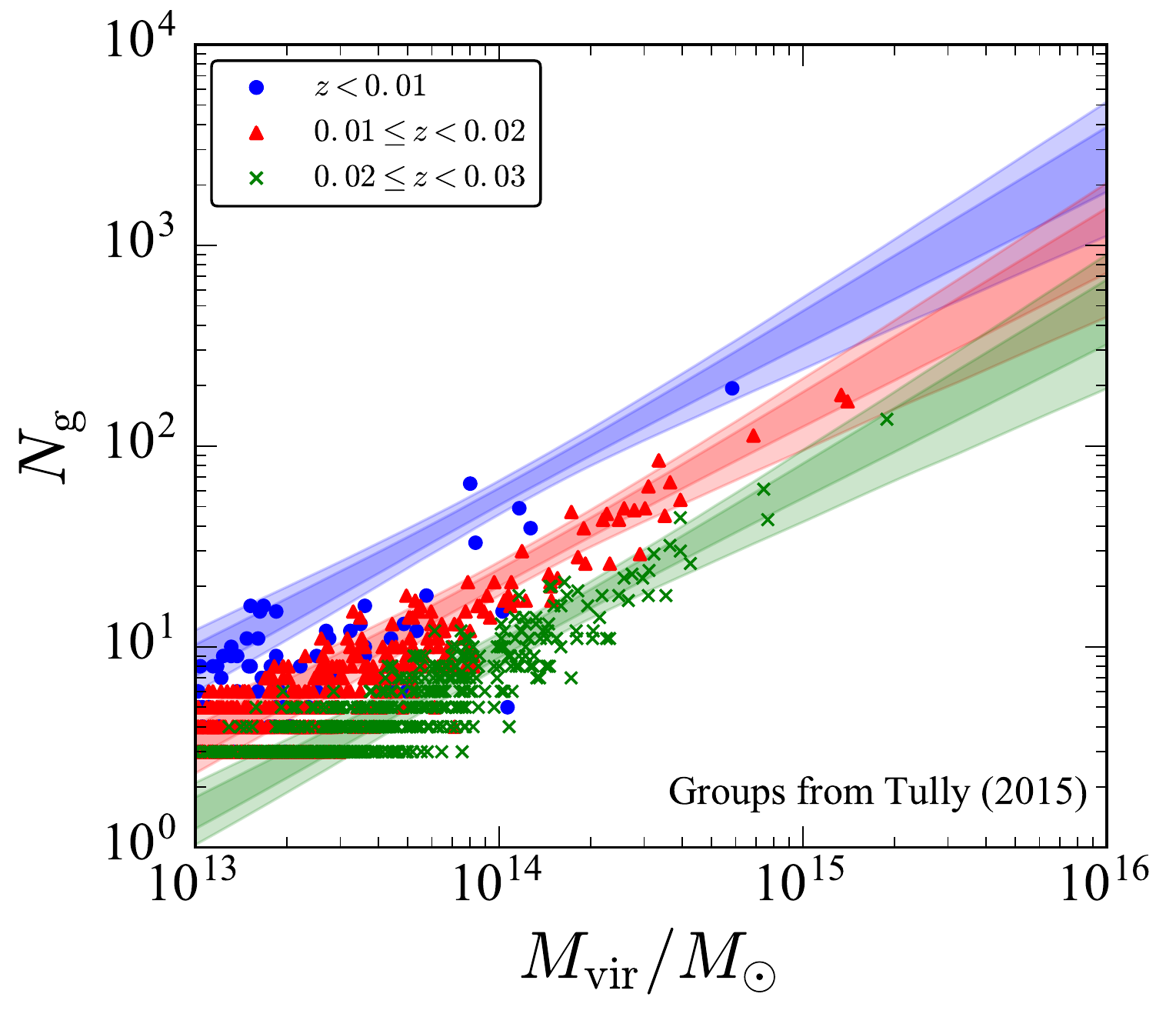}
  \includegraphics[width=8cm]{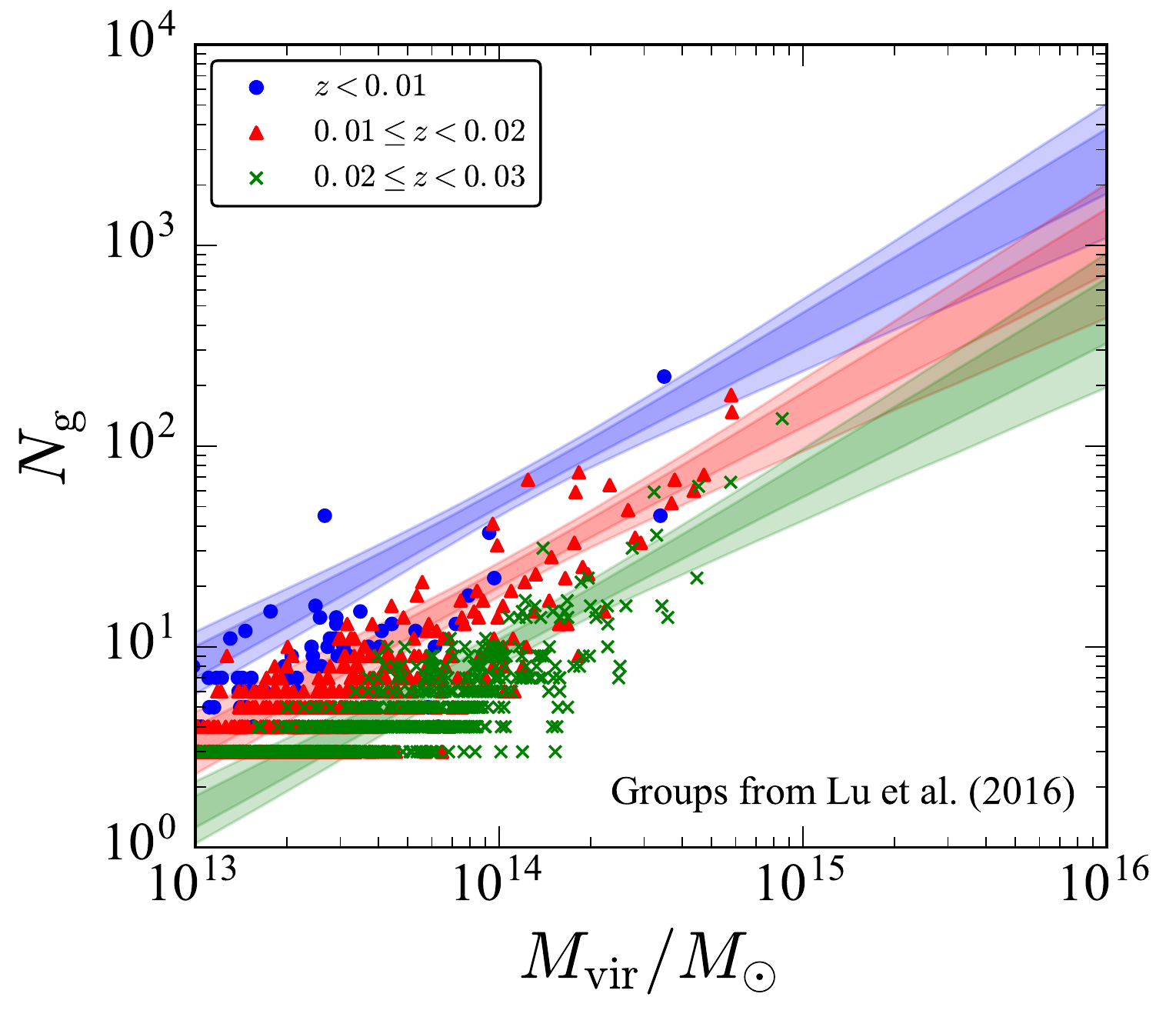}
  \caption{Numbers of member galaxies in groups and clusters of galaxies
  in the \citet[][top]{Tully2015} and \citet[][bottom]{Lu2016}
  catalogs. Different symbols show different redshift
  ranges. The thick and thin bands are the 68\% and 95\% credible intervals for
  the mean galaxy number predicted by the HOD model, $\langle N_{\rm
  g}|M\rangle$, after the selection effect has been
  corrected for each of these three subsamples (see text).}
  \label{fig:NgalHOD}
 \end{center}
\end{figure}

In the top panel of figure~\ref{fig:NgalHOD}, we show the numbers of
member galaxies
found in each group or cluster in the \citet{Tully2015} catalog as a
function of its virial mass $M_{\rm vir}$.
We estimate the virial mass of the groups and clusters through equation~(4)
of \citet{Tully2015b} and a proper conversion from $M_{200}$ to $M_{\rm
vir}$ afterwards.
We also show the 68\% and 95\% credible intervals for the mean galaxy
number, $\langle N_{\rm g}|M\rangle \equiv \langle N_{\rm cen}|M\rangle +
\langle N_{\rm sat}|M\rangle$, from the posterior distribution of the HOD
parameters of the ``run 1''.
The bottom panel of figure~\ref{fig:NgalHOD} shows the same, but for an
independent 2MRS group catalog by \citet{Lu2016}.

The HOD parameters are intrinsic to the galaxies. To
compare with the observation, we correct for the selection
effect as follows.
The number density of the {\it observed} galaxies is estimated from the
redshift distribution [equation~(\ref{eq:dNdz})] as $n_{\rm 2MRS}(z) =
(dN_{\rm g}/dz)(d\chi/dz)^{-1}/(4\pi f_{\rm sky} \chi^2)$, while the
{\it intrinsic} density is given by $n_{\rm HOD}(z) = \langle n_{\rm
g}(z)\rangle$ from equation~(\ref{eq:nHOD}).
In figure~\ref{fig:NgalHOD}, for each of the three subsamples (with mean
redshifts of $\bar z = 0.0065, 0.016$, and 0.025 for \citet{Tully2015}
and $\bar z = 0.0067, 0.016$, and 0.025 for \citet{Lu2016}), we re-scale
the total number of member galaxies $\langle N_{\rm g}|M\rangle$ by
$n_{\rm 2MRS}(\bar z) / n_{\rm HOD}(\bar z)$.
We see very good agreement between the HOD model and the observation of the
groups and clusters, justifying our model.

\section{Conclusion}
We have presented the measurement and interpretation of the angular power
spectrum of the 2MRS. We found that the 2MRS galaxies are so highly
clustered that the power spectrum at $\ell\gtrsim 30$ is dominated by
the one-halo term, i.e., galaxies residing in nearby galaxy groups and
clusters. We demonstrated this by the HOD modeling of the power
spectrum, as well as by directly summing up the contributions from
groups and clusters identified in the 2MRS catalog.

This property makes the 2MRS catalog well suited for cross-correlation
studies with other tracers of groups and clusters, e.g., hot gas traced
by the thermal Sunyaev-Zeldovich effect \citep{makiya}, gamma-rays
from annihilation of dark matter particles in sub-halos of groups and
clusters \citep{Ando2014}, and so on. Properly interpreting such
cross-correlation measurements requires accurate information of
clustering statistics of the 2MRS galaxies, and our measurement and HOD
model provide the required information. 

\section*{Acknowledgments}
We thank Lucas Macri for providing us with the 2MRS catalog.
This work was supported by the Netherlands Organization for Scientific
Research (NWO) through Vidi grant (SA), and also in part by JSPS KAKENHI
Grants, JP15H05896 (EK) and JP17H04836 (SA). We acknowledge the use
of HEALPix \citep{Gorski2005} and PolSpice \citep{spice} packages.

\end{document}